\documentclass[12pt]{article}

\usepackage{epsfig,bbold}

\catcode`@=12
\begin{document}

\thispagestyle{empty}
\begin{flushright} 
UCRHEP-T378\\ 
September 15, 2004
\end{flushright}
\vspace{0.5in}

\begin{center}
{\LARGE	\bf Quaternion Family Symmetry\\ of Quarks and
Leptons\\}
\vspace{1.5in}
{\bf Michele Frigerio$^1$, Satoru Kaneko$^2$, Ernest Ma$^1$, and Morimitsu 
Tanimoto$^3$\\}
\vspace{0.2in}
{\sl $^1$ Physics Department, University of California, Riverside, 
California 92521, USA\\}
{\sl $^2$ Physics Department, Ochanomizu University, Tokyo 112-8610, Japan\\}
{\sl $^3$ Physics Department, Niigata University, Niigata 950-2181, Japan\\}
\vspace{1.0in}
\end{center}

\begin{abstract}
To a first approximation, the quark mixing matrix has $\theta^q_{13} = 
\theta^q_{23} = 0$, whereas the lepton mixing matrix has $\theta^l_{23} = 
\pi/4$.  We show how this structure may be understood if the family symmetry
is $Q_8$, the quaternion group of eight elements.  
We find three viable scenarios for the Majorana neutrino mass matrix, 
each depending on 4 parameters and predicting a specific mass spectrum.  
The phenomenology of the two Higgs doublets which generate the Yukawa sector 
is analyzed and testable predictions are derived.
We discuss also the closely related model based on $D_4$, the symmetry group 
of the square. 
\end{abstract}

\begin{center}
\end{center}

\newpage
There are 3 families of quarks and leptons with accompanying $3 \times 3$ 
mixing matrices:  $V_{CKM} = V_u^\dagger V_d$ linking the $(u,c,t)$ quarks 
to the $(d,s,b)$ quarks, and $U_{MNSP} = U_l^\dagger U_\nu$ linking the 
$(e,\mu,\tau)$ charged leptons to the $(\nu_1,\nu_2,\nu_3)$ neutrinos. 
The $V_{CKM}$ matrix may be parametrized with 3 angles $\theta_{ij}^q$ 
and 1 phase $\delta^q$ 
and similarly $U_{MNSP}$, after absorbing two relative Majorana phases 
in the neutrino mass eigenvalues.  Numerically, $|V_{us}| \simeq 0.22$, 
$|V_{cb}| \simeq 0.04$, $|V_{ub}| \simeq 0.004$, thus setting $\theta^q_{13} 
= \theta^q_{23} = 0$ is a good first approximation of $V_{CKM}$.  On the 
other hand, $U_{MNSP}$ has a very different structure: 
both $\theta^l_{12}$ and $\theta^l_{23}$ are known to be large, with 
$\theta^l_{23} = \pi/4$ as its best value experimentally (and $\theta^l_{12} 
\neq \pi/4$).

We propose in this letter a new understanding of $V_{CKM}$ and $U_{MNSP}$ 
in terms of the non-Abelian discrete symmetry $Q_8$, the quaternion group of 
8 elements.  Using a natural generic 
assignment of quarks and leptons and the simplest nontrivial Higgs content, 
we show that there are only 4 possible scenarios for leptons (but only 3 are 
viable phenomenologically) and 1 scenario 
for quarks.  We will also discuss the closely related discrete symmetry 
$D_4$, which is the symmetry group of the square.  The latter also has 8 
elements and the same character table as $Q_8$ with 5 conjugacy classes and 5 
irreducible representations (irreps): ${\bf 1^{++}}$, ${\bf 1^{+-}}$, 
${\bf 1^{-+}}$, ${\bf 1^{--}}$, and ${\bf 2}$ (see Table \ref{char}). 
Two specific models based on $D_4 \times Z_2$ have recently been proposed 
\cite{gl,gjklt}.

\begin{table}[hbt]
\caption{Character table of $Q_8$ ($D_4$). Here $n$ is the number of 
elements in each conjugacy class, while $h$ is the order of any element $g$ 
in that class, i.e. the smallest integer such that $g^h = 1$.
}
\begin{center}
\begin{tabular}{|c|c|c|c|r|r|r|r|}
\hline
class & $n$ & $h$ & $\chi_{++}$ & $\chi_{+-}$ & $\chi_{-+}$ & 
$\chi_{--}$ & $\chi_{2}$ \\ 
\hline
$C_1$ & 1 & 1 & 1 & 1 & 1 & 1 & 2 \\ 
$C_2$ & 1 & 2 & 1 & 1 & 1 & 1 & $-2$ \\ 
$C_3$ & 2 & 4 & 1 & $-1$ & $-1$ & 1 & 0 \\ 
$C_4$ & 2 & 4(2) & 1 & 1 & $-1$ & $-1$ & 0 \\ 
$C_5$ & 2 & 4(2) & 1 & $-1$ & 1 & $-1$ & 0 \\
\hline
\end{tabular}
\end{center}
\label{char}\end{table}

Before we show how it all works, let us present our main results.  Using 
$Q_8$, we find the following 4 scenarios for the Majorana 
neutrino mass matrix (in the basis where the charged-lepton mass matrix 
is diagonal):
\begin{eqnarray}
{\cal M}^{(e,\mu,\tau)}_\nu = \pmatrix{a & c & d \cr c & 0 & b 
\cr d & b & 0},\label{one}\\
{\cal M}^{(e,\mu,\tau)}_\nu = \pmatrix{a & c & d \cr c & b 
& 0 \cr d & 0 & b},\label{two}\\
{\cal M}^{(e,\mu,\tau)}_\nu = \pmatrix{0 & c & d \cr c & a & b 
\cr d & b & a },\label{three}\\
{\cal M}^{(e,\mu,\tau)}_\nu = \pmatrix{0 & c & d \cr c & a 
& 0 \cr d & 0 & b}.\label{four}
\end{eqnarray}

\noindent 
Scenarios (1) and (4) correspond to mass matrices with two texture zeros,
which have been studied in the literature \cite{twozero}; in particular
scenario (4) is known to be ruled out. 
Scenarios (2) and (3) both have 
$m_{22} = m_{33}$ and one texture zero.  These are new structures which are 
also viable phenomenologically as we will show.

The group $Q_8$ may be generated by the eight $2 \times 2$ 
matrices $\pm {\mathbb 1}$, $\pm i \sigma_1$, $\pm i \sigma_2$, 
$\pm i \sigma_3$, whereas $D_4$ may be obtained with $\pm i \sigma_{1,3}$ 
replaced by $\pm \sigma_{1,3}$.  Each set is also a faithful two-dimensional 
irrep ${\bf 2}$ of the respective group.  Geometrically, the group 
$Q_8$ may be associated with the 8 vertices of the hyperoctahedron 
(dual of the hypercube) in four dimensions.
The two superscript signs for the one-dimensional irreps correspond 
to the characters of the $C_4$ and $C_5$ classes, 
generated by $\pm i \sigma_{1,3}$ (or 
$\pm \sigma_{1,3}$) respectively. Among themselves, the 4 one-dimensional 
representations transform as $Z_2 \times Z'_2$.
In the case of $Q_8$, the irreps ${\bf 1^{+-}},~{\bf 1^{-+}}$, and 
${\bf 1^{--}}$ are completely interchangeable, since they share exactly the
same group properties.  This means that if a theory contains a set of these 
irreps, replacing them with those obtained by any ($S_3$) permutation of the 3 
classes $C_{3,4,5}$ will not change the physical predictions of the theory. 
Specific examples will be given below.   In the case of $D_4$, only
${\bf 1^{+-}}$ and ${\bf 1^{-+}}$ are equivalent, since the conjugacy class
$C_3$ is distinguished from $C_4$ and $C_5$ by the order of their elements
(see Table \ref{char}).
For both groups, the basic tensor product rule is 
${\bf 2} \times {\bf 2} = {\bf 1^{++}} + {\bf 1^{+-}} + {\bf 1^{-+}} + 
{\bf 1^{--}}$, but the doublet components are combined in different ways 
for $Q_8$ and $D_4$, as shown in Table \ref{deco}. 
Since the two-dimensional irrep of $D_4$ is real, 
$(\psi_1, \psi_2)$ transforming as a doublet implies that 
$(\psi^*_1, \psi^*_2)$ is also a doublet, whereas in the case of $Q_8$, the 
correct assignment is $(\psi^*_2, -\psi^*_1)$
as expected, just like a doublet under $SU(2)$.  

\begin{table}[htb]
\caption{${\bf 2} \times {\bf 2}$ decompositions of $Q_8$ and 
$D_4$.}
\begin{center}
\begin{tabular}{|c|c|c|c|c|}
\hline
group & $11+22$ & $12+21$ & $11-22$ & $12-21$ \\ 
\hline
$Q_8$ & ${\bf 1^{--}}$ & ${\bf 1^{-+}}$ & ${\bf 1^{+-}}$ & ${\bf 1^{++}}$ \\ 
$D_4$ & ${\bf 1^{++}}$ & ${\bf 1^{+-}}$ & ${\bf 1^{-+}}$ & ${\bf 1^{--}}$ \\ 
\hline
\end{tabular}
\end{center}
\label{deco}
\end{table}

Under $Q_8$, we assign the 3 quark and lepton families and two Higgs
doublets as follows:
\begin{eqnarray}
&&(u_i~d_i), ~u^c_i, ~d^c_i \sim {\bf 1^{--}}, ~{\bf 1^{-+}}, 
~{\bf 1^{+-}}~; \label{qrep}\\ 
&&(\nu_i~l_i), ~l^c_i \sim {\bf 1^{++}}, ~{\bf 2}~;\\
&&(\phi_1^0, \phi_1^-) \sim {\bf 1^{++}}, ~~~(\phi_2^0, \phi_2^-) 
\sim {\bf 1^{+-}}~.
\label{doublets}
\end{eqnarray}
As a result, each quark mass matrix in the basis $({\bf 1^{--}},
{\bf 1^{-+}},{\bf 1^{+-}})$ is of the form
\begin{equation}
{\cal M}_q = \pmatrix{ a & d & 0 \cr e & b & 0 \cr 0 & 0 & c
},
\label{Mq}
\end{equation}
where $a,b,c$ are proportional to $\langle \phi_1^0 \rangle$ and $d,e$ to 
$\langle \phi_2^0 \rangle$.  This means that the third family does not 
mix with the other two, i.e. $\theta^q_{13} = \theta^q_{23} = 0$, which 
is a good first approximation.  On the other hand, 
the charged-lepton mass matrix in the basis $({\bf 1^{++}},{\bf 2})$ is 
given by
\begin{equation}
{\cal M}_l = \pmatrix{ a & 0 & 0 \cr 0 & c & b \cr 0 & -b & -c
},
\label{Ml}
\end{equation}
where $a,b$ are proportional to $\langle \phi_1^0 \rangle$ and $c$ to 
$\langle \phi_2^0 \rangle$.  This matrix is easily diagonalized by a 
rotation of $\pi/4$ on the left and on the right, i.e. $\mu,\tau = (l_2 
\pm l_3)/\sqrt 2$ and $\mu^c,\tau^c = (l_2^c \mp l_3^c)/\sqrt 2$, with $m_e 
= |a|$, $m_\mu = |c-b|$, and $m_\tau = |c+b|$.  

The neutrino mass matrix is assumed to be Majorana and generated by the 
naturally small vacuum expectation values (VEVs) of heavy Higgs triplets 
$\xi_i \equiv (\xi_i^{++}, \xi_i^+, \xi^0_i)$. The assignment of triplets
to the different $Q_8$ irreps is crucial for the resulting neutrino mass 
pattern.   

\begin{figure}[htb]
\begin{center}\includegraphics[width=260pt]{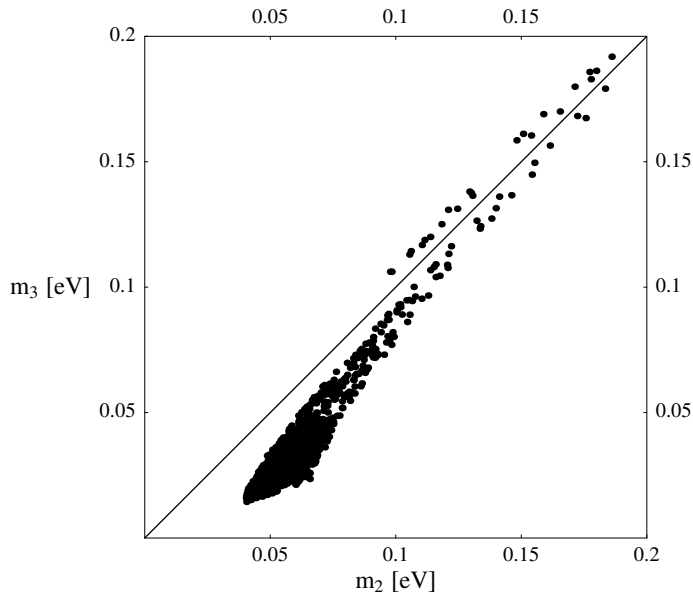}
\end{center}
\caption{The allowed region in $m_2 - m_3$ plane for scenario (1) is 
presented in eV units. The masses are scanned in the experimental allowed 
range : $\Delta m_{\rm sol}^2=(7.7 - 8.8)\times 10^{-5}$ eV$^2$, 
$\Delta m_{\rm atm}^2=(1.5 - 3.4)\times 10^{-3}$ eV$^2$, 
$\tan^2 \theta_{\rm sol}=0.33 - 0.49$, $\sin^2 2\theta_{\rm atm}\geq 0.92$, 
and $\sin\theta_{\rm CHOOZ}<0.2$.}
\label{s1}
\end{figure}

{\bf Scenario (1).}~
Since the triplet VEVs are induced \cite{ms98} via trilinear 
couplings of the form $\xi_i\phi_j\phi_k$, the requirement of $Q_8$ symmetry 
would allow only $\xi_1\sim{\bf 1^{++}}$ and $\xi_2\sim{\bf 1^{+-}}$ to 
contribute. In that case, ${\cal M}_\nu$ has $|m_2| = |m_3|$ and $\nu_e$ is 
unmixed, which is not 
realistic. However, $Q_8$ is expected to be broken softly, thus the scalar 
trilinear $\xi_i\phi_j\phi_k$ terms may induce small VEVs also on 
$(\xi_3, \xi_4) \sim {\bf 2}$, by which
\begin{equation}
{\cal M}_\nu = \pmatrix{ a & e & f \cr e & b & 0 \cr f & 0 & -b
}~,
\label{1mn}
\end{equation}
where $a$ comes from $\langle \xi_1^0 \rangle$, $b$ from $\langle \xi_2^0 
\rangle$, $e = h \langle \xi^0_4 \rangle$, and $f = -h \langle \xi^0_3 
\rangle$.  In the $e, \mu, \tau$ basis with $\mu,\tau = (l_2 \pm l_3)/
\sqrt 2$, the neutrino mass matrix takes the form of Eq.~(\ref{one}),
with $c,d = (e \pm f)/\sqrt 2$.
Two texture zeros in the $\nu_\mu \nu_\mu$ and $\nu_\tau \nu_\tau$ 
entries are thus derived (for the first time) by our 
application of the $Q_8$ family symmetry.  This is known to be a viable 
pattern \cite{twozero,fs03} and in most cases 
predicts an inverted mass spectrum.  
Here $\theta^l_{23} = \pi/4$ and $\theta^l_{13} = 0$ are obtained in the 
limit of $c = \pm d$.  
Deviations from maximal $2-3$ mixing are allowed
proportionally to the nonzero value of $\theta_{13}^l$, which can be 
as large as the experimental upper bound.
With the present experimental constraints, we find 
$|m_2| > 0.04$ eV and $|m_3| > 0.015$ eV, as shown in Fig.~1.
The neutrinoless $2\beta$-decay rate is controlled by 
$m_{ee}\equiv |a| > 0.02$ eV.
A quasi-degenerate spectrum can be obtained, when in Eq.(\ref{one}) 
$a\approx b$ and $c,d$ are much smaller \cite{fs03}; 
in this limit the ordering of the
spectrum can be also normal (see Fig.~\ref{s1}).

\begin{table}[htb]
\caption{Assignments of the Higgs doublet $\phi_2$ and of the
Higgs triplets $\xi_1$ and $\xi_2$ corresponding to
the four scenarios for the neutrino mass matrix described by 
Eqs.~(\ref{one})-(\ref{four}). The assignments $\phi_1\sim{\bf 1^{++}}$ and
$(\xi_3,\xi_4)\sim{\bf 2}$ are kept fixed in all cases under study.}
\begin{center}
\begin{tabular}{|c|cc|c|c|c|}
\hline
&&& 1 & 2 & 3 \\
\hline
& $\xi_1$ & $\xi_2$ & $\phi_2\sim{\bf 1^{+-}}$ & 
$\phi_2\sim{\bf 1^{-+}}$ & $\phi_2\sim{\bf 1^{--}}$ \\
\hline 
A & ${\bf 1^{++}}$ & ${\bf 1^{+-}}$ & (1) & (2) & (2) \\ 
B & ${\bf 1^{++}}$ & ${\bf 1^{-+}}$ & (2) & (1) & (2) \\ 
C & ${\bf 1^{++}}$ & ${\bf 1^{--}}$ & (2) & (2) & (1) \\ 
D & ${\bf 1^{+-}}$ & ${\bf 1^{-+}}$ & (3) & (3) & (4) \\ 
E & ${\bf 1^{+-}}$ & ${\bf 1^{--}}$ & (3) & (4) & (3) \\ 
F & ${\bf 1^{-+}}$ & ${\bf 1^{--}}$ & (4) & (3) & (3) \\ 
\hline
\end{tabular}
\end{center}
\label{higgs}
\end{table}

There are 3 equivalent assignments of Higgs doublets and triplets which 
result in scenario (1), as listed in Table \ref{higgs}.  In the choice B2,
${\cal M}_l$ is diagonal if we redefine $l^c_{2,3}$ as $l^c_{3,2}$, and 
${\cal M}_\nu$ has the form of Eq.~(\ref{one}) automatically as it should. 
In the choice C3, the charged-lepton and neutrino mass matrices 
are given by
\begin{equation}
{\cal M}_l = \pmatrix{a & 0 & 0 \cr 0 & c & b \cr 0 & -b & c
}~, ~~~ 
{\cal M}_\nu = \pmatrix{a & e & f \cr e & b & 0 \cr f & 0 & b
}~.
\label{Ml2}
\end{equation}
Diagonalizing ${\cal M}_l$ 
by
\begin{equation}
U = \pmatrix{1 & 0 & 0 \cr 0 & -i/\sqrt 2 & 1/\sqrt 2 \cr 0 & 
i/\sqrt 2 & 1/\sqrt 2}~,
\label{U2}
\end{equation}
we find that ${\cal M}^{(e,\mu,\tau)}_\nu = U {\cal M}_\nu U^T$ 
has the same form as
Eq.~(\ref{one}), with $c,d = (f \pm ie)/\sqrt 2$. 
In the quark sector, if we replace the 3 one-dimensional irreps of 
Eq.~(\ref{qrep}) 
with any other 3, and put the remaining one in the lepton sector, 
we again have the same physical predictions, i.e. Eqs.~(\ref{Mq}) 
and (\ref{one}).

{\bf Scenario (2).}~
Instead of using 
Higgs triplets of the same one-dimensional irreps as the Higgs doublets which 
we did 
in scenario (1), consider the use of one triplet in the  
${\bf 1^{++}}$ irrep (as $\phi_1$)
and a second triplet in one irrep equivalent but different from that of 
$\phi_2$. There are 6 such assignments, as shown in Table \ref{higgs}, which
are all equivalent as expected. 
For definiteness, let us study the choice B1.  
In this case, ${\cal M}_l$ is given by Eq.~(\ref{Ml})  
and ${\cal M}_\nu$ by Eq.~(\ref{one}), 
so that in the $e,\mu,\tau$ basis we obtain Eq.~(\ref{two})
(with a redefinition of $c$ and $d$).
Let us now prove that this new pattern is viable phenomenologically. 
If $c$ and $d$ are relatively real, than 
$m_3 = b$ and $\nu_3 = s_{23}\nu_\mu + c_{23}\nu_\tau$, 
with $s_{23} = -d/\sqrt{c^2+d^2}$, 
and $c_{23} = c/\sqrt{c^2+d^2}$, i.e. $\theta^l_{13} = 0$ necessarily and 
$\tan\theta^l_{23}=-d/c$ is arbitrary  
(the second $D_4 \times Z_2$ model \cite{gjklt} 
also has this property, but its other predictions 
are different). In this limiting case we have 
the identity $a = m_1 + m_2 - m_3$ and the sum rule
$s_{12}^2 m_1 + c_{12}^2 m_2 = m_3$,
which implies inverted hierarchy as shown in Fig.~2.

\begin{figure}[htb]
\begin{center}\includegraphics[width=260pt]{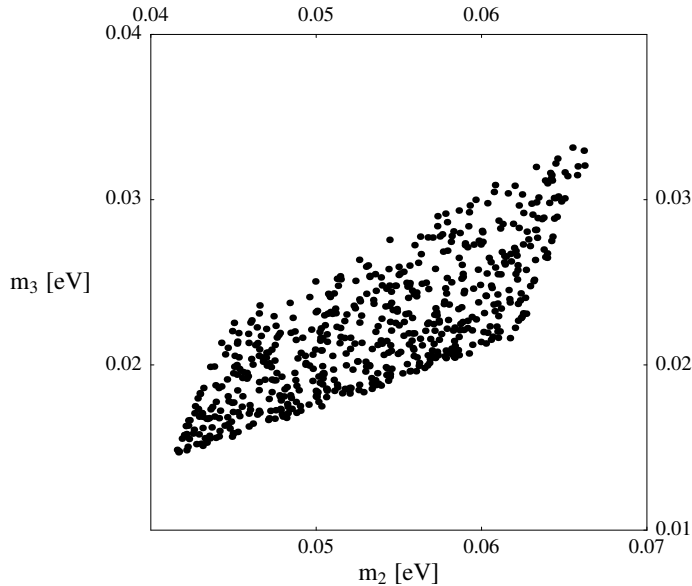}\end{center}
\caption{The allowed region in $m_2 - m_3$ plane for 
scenario (2) is presented in eV units,  in the limit 
$\theta^l_{13}=0$. The masses are scanned in the same 
experimental allowed range of neutrino parameters as Fig.~\ref{s1}.}
\label{s2}
\end{figure}

If $c$ and $d$ are not relatively real, $\theta^l_{13}$ becomes nonzero, 
and the resulting allowed region in the $m_2 - m_3$ plane 
is similar to that of Fig.~\ref{s1}. In particular, a degenerate spectrum 
is allowed, when the matrix (\ref{two}) becomes close to the identity matrix
($a\approx b$ and $c,d$ much smaller) \cite{fs03}.

{\bf Scenario (3).}~ 
The last viable scenario under $Q_8$ has one Higgs triplet 
in the same irrep as $\phi_2$ and a second triplet in one of the other two 
equivalent irreps.  There are 6 equivalent assignments, as listed in Table 
\ref{higgs}. Consider the choice E1.  
In this case, ${\cal M}_l$ is given by Eq.~(\ref{Ml}) as 
before, but the neutrino mass matrix is now
\begin{equation}
{\cal M}_\nu = \pmatrix{ 0 & e & f \cr e & a+b & 0 \cr f & 0 & a-b
}.
\label{Mn}
\end{equation}
Since this is the same as Eq.~(\ref{1mn}) except for the placement of $a$,  
it becomes the matrix of Eq.~(\ref{three}) in the $e,\mu,\tau$ basis, 
in analogy with Eq.~(\ref{one}).
This is another new viable pattern 
and predicts a normal hierarchy.  Here $\theta^l_{23} = \pi/4$ and 
$\theta^l_{13} = 0$ are again obtained in the limit $c=\pm d$. 
With the form of Eq.~(\ref{three}), the constraint $\sin \theta_{13} < 0.2$ 
implies that $\sin^2 2 \theta_{23} > 0.987$, and predicts
\begin{equation}\begin{array}{c}
0.035~{\rm eV} < |m_3| < 0.065~{\rm eV}~,\\ 0.009~{\rm eV} < |m_2| < 
0.015~{\rm eV}.
\label{range}
\end{array}\end{equation}

Consider the choice F3. In this case, ${\cal M}_l$ is the one given in 
Eq.~(\ref{Ml2}), which is diagonalized by $U$ of 
Eq.~(\ref{U2}). 
On the other hand, the 
neutrino mass matrix is of the form of Eq.~(\ref{three}) already, 
so isn't the extra 
rotation of $\pi/4$ required by $U$ going to change its form?  The answer 
is no, because
\begin{equation}
\frac{1}{\sqrt 2} \pmatrix{-i & 1 \cr i & 1} 
\frac{1}{\sqrt 2} 
\pmatrix{1 & 1 \cr 1 & -1} 
= \frac{1}{2} \pmatrix{1-i & -1-i \cr 1+i & -1+i},
\end{equation}
which also rotates the $\nu_\mu-\nu_\tau$ sector by $\pi/4$.  As a result, 
Eq.~(\ref{three}) retains its form: the observed maximal atmospheric mixing  
emerges from maximal mixing in both neutrino and charged lepton sectors!
It is easy to work out the details in each of 
the other cases as well.

In all 3 scenarios, i.e. Eqs.~(\ref{one}), (\ref{two}), and (\ref{three}), 
there are 4 parameters in 
${\cal M}^{(e,\mu,\tau)}_\nu$.  This means that the 3 neutrino masses and 
their 3 mixing angles are related. 
In particular, the absolute scale of $m_{1,2,3}$ is constrained in each 
case, as shown in Figs.~1, 2, and Eq.~(\ref{range}).  
Thus they are all testable predictions.

So far we have a successful $U_{MNSP}$ matrix but only an approximate 
$V_{CKM}$ matrix because $\theta^q_{13} = \theta^q_{23} = 0$.  This means 
that the quark sector has an extra global symmetry associated with the 
third family, i.e. a top/bottom flavor number.  This symmetry may be broken, 
for example, by adding to the specific assignment of Eq.~(\ref{doublets}) 
the Higgs doublet $\phi_3 \sim {\bf 1^{--}}$ which contributes to the (23) 
and (32) entries of ${\cal M}_q$ in Eq.~(\ref{Mq}) to allow $\theta^q_{23} 
\neq 0$ and also $\theta^q_{13} \neq 0$. 
This will of course also affect 
${\cal M}_l$ of Eq.~(\ref{Ml}) but only in the $\mu-\tau$ sector. 
It means that even if $\theta^l_{13} = 0$ in scenarios (1) and (3), 
$\theta^l_{23}$ should deviate from 
$\pi/4$ as well, having the same origin as deviations of 
$\theta^q_{23}$ and $\theta^q_{13}$ from zero.  (For a possible indication 
of deviation from maximal atmospheric mixing in SuperKamiokande data see 
for example Ref.~\cite{peres}.)

If we use $D_4$ instead of $Q_8$, then according to the multiplication 
rules of Table 2, we find one realization of scenario (1), i.e. the choice 
C3 of Table \ref{higgs}, two of scenario (2), i.e. the choices C1
and C2, and two of scenario (3), i.e. 
the choices D1 and D2. There are also two other viable scenarios (but 
with 5 parameters), i.e.
$$
{\cal M}^{(e,\mu,\tau)}_\nu = \pmatrix{e & c & d \cr c & a & b 
\cr d & b & a} ~(3');~
{\cal M}^{(e,\mu,\tau)}_\nu = \pmatrix{e & c & d \cr c & a & 0 
\cr d & 0 & b} ~(4'). 
$$
Scenario $(3')$ is obtained with the choices A2,A3,B1,B3 of 
Table \ref{higgs} and 
scenario $(4')$ with A1,B2.  If we assume $c=d$ in scenario $(3')$, 
then we obtain the result of the first $D_4 \times Z_2$ model 
\cite{gl}, but without using the extra $Z_2$.

In the Higgs sector, $\phi_1$ and $\phi_2$ are distinguished by an odd-even 
parity.  The ensuing scalar potential is well-known, having in general 
a minimum with nonzero VEVs for both $\phi_1^0$ and $\phi_2^0$. 
Consider ${\cal M}_q$ of Eq.~(\ref{Mq}).  
The $2 \times 2$ sub-matrix spanning the 
first two families may be diagonalized in general by a rotation on 
the left and a rotation on the right.  Specifically, in the case of the 
$d$ and $s$ quarks, it may be written as
\begin{equation}\begin{array}{c}
\pmatrix{c_L & s_L \cr -s_L & c_L} 
\pmatrix{m_d & 0 \cr 0 & m_s }
\pmatrix{c_R & -s_R \cr s_R & c_R }
= \\  \pmatrix{c_L c_R m_d + s_L s_R m_s & 
-c_L s_R m_d + s_L c_R m_s \cr -s_L c_R m_d + c_L s_R m_s & s_L s_R m_d + 
c_L c_R m_s}.
\end{array}\end{equation}
Since $\phi_1^0$ couples to the diagonal entries and $\phi_2^0$ couples to 
the off-diagonal entries, there are $d s^c$ and $s d^c$ couplings to each 
given by
\begin{eqnarray}
{\cal L}_Y &\supset&  
\left\{ [ -s_R c_R (c_L^2 - s_L^2) m_d + s_L c_L (c_R^2 - s_R^2) m_s]~d s^c 
\right. \nonumber \\ & + & \left. [ -s_L c_L (c_R^2 - s_R^2) m_d + s_R c_R 
(c_L^2 - s_L^2) m_s]~s d^c \right\} \nonumber \\ && 
\times\left( \phi_2^0/v_2 - \phi_1^0/v_1 \right)+ H.c.
\label{FCNC}
\end{eqnarray} 
Note that, even though ${\cal M}_l$ of Eq.~(\ref{Ml}) is not diagonal, 
the $\mu-\tau$ sector has $c_L^2 = s_L^2 = c_R^2 = s_R^2 = 1/2$.  
Hence Eq.~(\ref{FCNC}) shows that 
flavor-changing $\mu-\tau$ interactions are absent.

In our model, the state $(v_1 \phi_1^0 + v_2 \phi_2^0)/\sqrt{v_1^2+v_2^2}$ 
is identifiable with the neutral Higgs boson of the Standard Model.  Its 
orthogonal state $h^0 = (v_1 \phi_2^0 - v_2 \phi_1^0)/\sqrt{v_1^2+v_2^2}$ 
appears in Eq.~(\ref{FCNC}) and couples to both $d s^c$ and $s d^c$, thereby 
contributing to the $K_L-K_S$ mass difference.  This contribution depends 
on the unknown parameters $s_R$ and $m_h$.  It is suitably suppressed if 
either $s_R$ is small or $m_h$ is large.  For example, if $s_R$ is 
negligible, then \cite{ma01}
\begin{equation}
\frac{\Delta m_K}{m_K} \simeq \frac{B_K f_K^2}{3 m_h^2} 
\left( \frac{v_1^2+v_2^2} 
{v_1^2 v_2^2} \right) s_L^2 c_L^2 m_d m_s.
\end{equation}
Taking $v_1=v_2=123$ GeV, $s_L^2 \simeq m_d/m_s$, 
$B_K = 0.4$, $f_K = 114$ MeV and $m_d = 7$ MeV, this contribution 
is $1.1 \times 10^{-15} (100~{\rm GeV}/m_h)^2$, the 
experimental value being $7.0 \times 10^{-15}$.  Similarly, the contribution 
to $\Delta m_D/m_D$ is estimated to be $1.3 \times 10^{-15} (100~{\rm GeV}
/m_h)^2$, which is well below the experimental upper bound of $2.5 \times 
10^{-14}$.
In the leptonic sector, $h^0$ is predicted to have the interaction
\begin{equation}\begin{array}{c}
{\cal L}_Y \supset 
\frac{h^0}{2 \sqrt {v_1^2+v_2^2}} \left\{ \left( \frac{v_1}{v_2} 
- \frac{v_2 }{v_1} \right) (m_\tau \tau \tau^c + m_\mu \mu \mu^c) \right.+ 
\\ +\left.
\left( \frac{v_1}{v_2} + \frac{v_2}{v_1} \right) (m_\mu \tau \tau^c + 
m_\tau \mu \mu^c) \right\} + H.c.
\end{array}\end{equation}
Its decay into $\tau^+ \tau^-$ and $\mu^+ \mu^-$ pairs will 
be crucial in the verification of this model.

{\bf Acknowledgments.}
Three of us (S.K., E.M., M.T.) thank the Yukawa Institute for Theoretical 
Physics at Kyoto University for support at the Workshop YITP-W-04-08 
(Summer Institute 2004, Fuji-Yoshida), where this research was initiated. 
This work was supported in part by the U.~S.~Department of Energy
under Grant No.~DE-FG03-94ER40837.

\bibliographystyle{unsrt}

\end{document}